\documentclass{jfm}

\usepackage{graphicx}
\usepackage{natbib}
\usepackage{amsmath}
\usepackage{esdiff}
\usepackage{float}
\usepackage{microtype}

\ifCUPmtlplainloaded \else
  \checkfont{eurm10}
  \iffontfound
    \IfFileExists{upmath.sty}
      {\typeout{^^JFound AMS Euler Roman fonts on the system,
                   using the 'upmath' package.^^J}%
       \usepackage{upmath}}
      {\typeout{^^JFound AMS Euler Roman fonts on the system, but you
                   dont seem to have the}%
       \typeout{'upmath' package installed. JFM.cls can take advantage
                 of these fonts,^^Jif you use 'upmath' package.^^J}%
      }
  \else
  \fi
\fi

\ifCUPmtlplainloaded \else
  \checkfont{msam10}
  \iffontfound
    \IfFileExists{amssymb.sty}
      {\typeout{^^JFound AMS Symbol fonts on the system, using the
                'amssymb' package.^^J}%
       \usepackage{amssymb}%

      }{}
  \fi
\fi

\ifCUPmtlplainloaded \else
  \IfFileExists{amsbsy.sty}
    {\typeout{^^JFound the 'amsbsy' package on the system, using it.^^J}%
     \usepackage{amsbsy}}
    {\providecommand\boldsymbol[1]{\mbox{\boldmath $##1$}}}
\fi

\newsavebox{\astrutbox}
\sbox{\astrutbox}{\rule[-5pt]{0pt}{20pt}}

\title{Interaction Between Two Closely-Spaced Waving Slender Elastic Cylinders Immersed in a Viscous Fluid}
\shorttitle{Interaction Between Two Waving Slender Elastic Cylinders}
\author[R. Elfasi, Y. Elimelech and A.D. Gat]{R. Elfasi$^1$, Y. Elimelech$^1$ and A.D. Gat$^{1,2}$}
\affiliation{$^1$Autonomous Systems and Robotics Department, Technion - Israel Institute of Technology, Haifa 3200003, Israel\\$^2$Faculty of Mechanical Engineering, Technion - Israel Institute of Technology, Haifa 3200003, Israel}

\pubyear{2016} \volume{999} \pagerange{999--9999}
\date{2016}
\begin{document}
\maketitle

\begin{abstract}
We study the hydrodynamic interaction between two closely-spaced waving elastic cylinders immersed within a viscous liquid, at the creeping flow regime. The cylinders are actuated by a forced oscillation of the slope at their clamped end and are free at the opposite end. We obtain an expression for the interaction force and apply an asymptotic expansion based on a small parameter representing the ratio between the elastic deflections and the distance between the cylinders. The leading-order solution is an asymmetric oscillation pattern at the two frequencies ($\omega_1,\omega_2$) in which the cylinders are actuated. Higher orders oscillate at frequencies which are combinations of the actuation frequencies, where the first-order includes the $2\omega_1,2\omega_2,\omega_1+\omega_2$, and $\omega_1-\omega_2$ harmonics. For in-phase actuation with $\omega_1= \omega_2$, the deflection dynamics are identical to an isolated cylinder with a modified Sperm number. For configurations with  $\omega_1\approx \omega_2$, the  $\omega_1-\omega_2$ mode represents the dominant first-order interaction effect due to significantly smaller effective Sperm number. Experiments are conducted to verify and illustrate the theoretical predictions. 
\end{abstract}

\section{Introduction}
We study the interaction between two closely-spaced oscillating elastic cylinders which are immersed within a viscous liquid. The cylinders are actuated by a forced oscillation of the slope at their clamped ends, which may vary in frequency, amplitude, and phase. We focus on configurations with negligible inertial effects and linear elasticity, where the dynamics are governed by a balance between viscous and elastic forces. 

Various previous works examined the viscous-elastic dynamics of a single elastic cylinder actuated by a forced oscillation at its clamped end. These include \cite{machin1958wave} who was the first to analytically obtain the deflection modes of such a passive elastic filament for the case of actuation of the slope at the fixed end. Using a similar approach, \cite{wiggins1998flexive} and \cite{wiggins1998trapping} studied deflection modes and  propulsion forces for forced oscillations and impulses of the position of the fixed end, combined with a requirement of zero torque. An experimental study was conducted by \cite{tony2006experimental}, who measured both deflection and propulsion for a single elastic filament actuated by oscillation of the slope. The experimental data showed good agreement with both linear and non-linear theoretical predictions. Other relevant works include \cite{camalet2000generic} who studied the dynamics of an elastic cylinder actuated by internal moments and \cite{arco2014viscous} who experimentally studied oscillating flexible sheet as a novel pumping mechanism in the creeping flow regime. 

Previous studies on interaction between two oscillating elastic cylinders focused mainly on forced deformations in the context of synchronization dynamics between closely-spaced waving flagella. One of the first works on synchronization of flagella was conduced by \cite{Taylor} who studied the simplified model of two infinite sheets with prescribed waveforms and showed that energy dissipation is minimized when the sheets oscillate in-phase. More recently, \cite{elfring2009hydrodynamic} analyzed a similar simplified configuration and concluded that synchronization may occur solely from hydrodynamics forces and requires front-back asymmetry of the deformation modes. Other works focused on experiments in biological systems, including \cite{brumley2014flagellar} who demonstrated that synchronization dynamics of the \textit{Volvox carteri} may indeed occur due to the hydrodynamic effects alone.

The aim of this work is to analytically and experimentally study the deflection dynamics of two interacting passive elastic cylinders actuated by a forced oscillation of the slope at their clamped end. This work is arranged as follows: In \S2 we present the problem formulation, compute the interaction forces and apply asymptotic expansions. In \S3 we present the deflection modes, define the experimental methodology and compare the experimental data to the analytical results. In \S4 we give concluding remarks.

\section{Analysis}
\begin{figure}
    \begin{center}
        \includegraphics[width=0.6\textwidth]{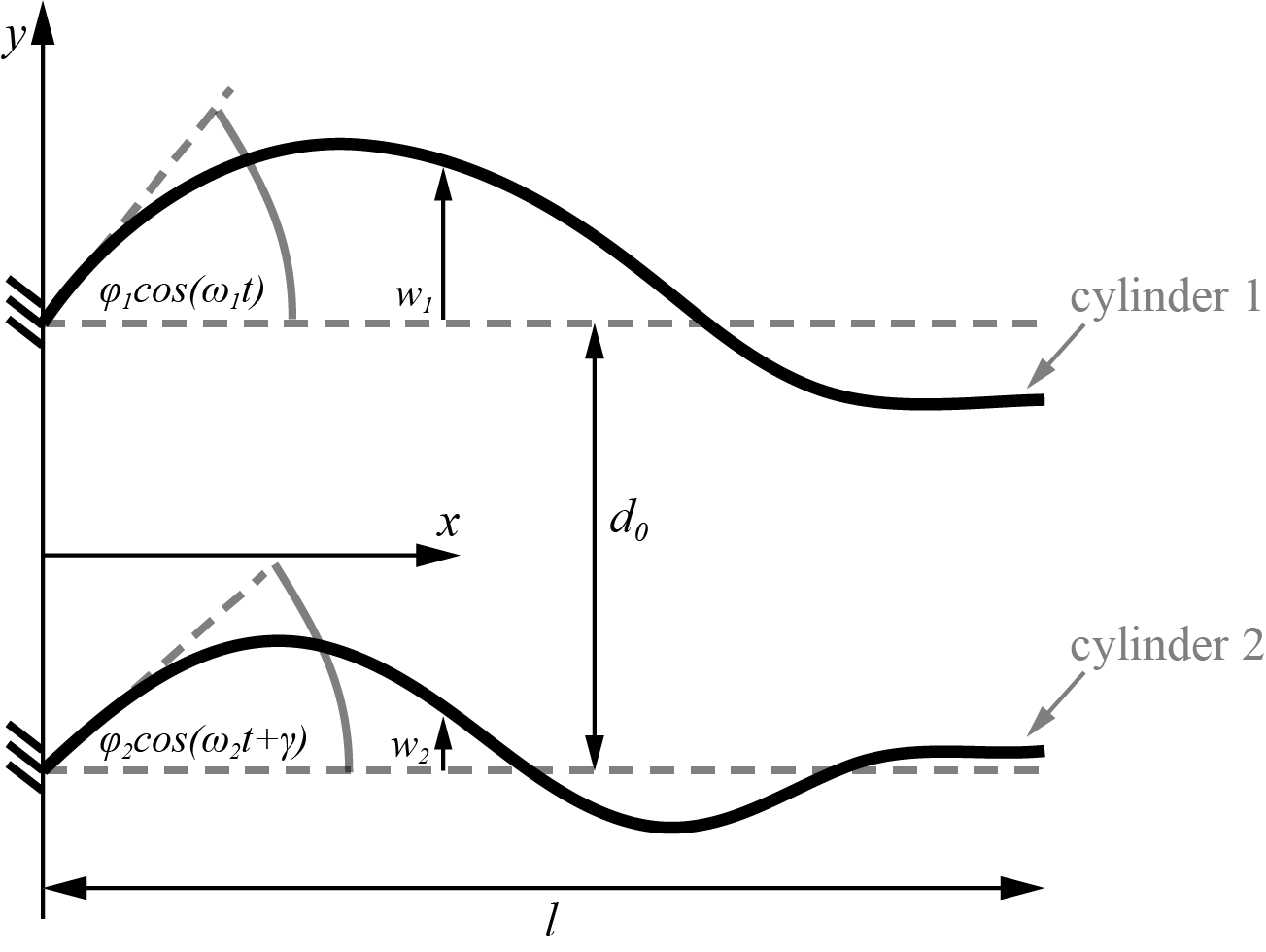}
        \caption{Illustration of the examined configuration consisting of two oscillating elastic cylinders immersed in a viscous fluid. The cylinders, at rest, are parallel to the $x$ axis and their centers oscillate within the $x-y$ plane. The distance between the bases of the cylinders is $d_0$ and the deflections are denoted by $w_1$ and $w_2$ for cylinder $1$ and $2$, respectively. The length of both cylinders is $l$.}
 \label{fig:new_problem}
    \end{center}
\end{figure}

We examine the fluidic interaction between two closely-spaced slender elastic cylinders immersed in a viscous liquid and actuated due to a forced oscillation of the slope at their clamped end. The coordinates and configuration are illustrated in Fig. \ref{fig:new_problem}. The Cartesian coordinate system is denoted by $(x,y,z)$ and time is denoted by $t$. The cylinders, at rest, are parallel to the $x$-direction and their centers oscillate within the $x-y$ plane. The fluid viscosity and density are denoted by $\mu$ and $\rho$, respectively. The cylinder flexural rigidity is $s$, the beam mass per-unit-length is $m$, the gap between the centers of the cylinders is $d$ and the gap at rest is $d_0$. The radius and length of the cylinders are $r_c$ and $l$, respectively. The forced oscillations of the slope of the cylinders at $x=0$ are at frequencies $\omega_i$ and amplitudes $\phi_i$ (where $i=1$ and $i=2$ denote cylinders $1$ and $2$, respectively). The phase difference between the forced oscillations is $\gamma$. The deflection of the cylinders is $w_i$,  where we define an auxiliary average deflection $w_a=(w_1+w_2)/2$ and an auxiliary relative deflection $w_d=(w_1-w_2)/2$. The perpendicular drag coefficient of the cylinders is $\xi_\perp$.

Hereafter, asterisk superscripts denote characteristic values and Capital letters denote normalized variables. The characteristic average deflection is $w_a^*$, characteristic relative deflection is $w_d^*$ and characteristic frequency is $\omega^*$. We define the small parameters (where $w^*=\max{(w^*_a,w^*_d)}$)
\begin{equation} \label{eq:2.1}
    \frac{\rho \omega^* (w^*)^2}{\mu}\ll1,\quad \frac{m (\omega^* w^*)^2}{s}\ll1,\quad 
    \frac{d_0}{l}\ll1, \quad \frac{r_c}{d_0}\ll1,\quad \frac{w_a^*}{l}\ll1, \quad \frac{w_d^*}{d_0}\ll1, 
\end{equation}
corresponding to assumptions of negligible fluidic inertia (small Womersley number), negligible solid inertia, small gap to cylinder length ratio, small cylinder radius to gap ratio, small average deflection to length ratio and small relative deflection to gap ratio.

In addition, we apply the commonly used approximation  \citep[see][]{machin1958wave,gray1955propulsion,lighthill1975mathematica,wiggins1998flexive,wiggins1998trapping,powers2002role, tony2006experimental,friedrich2010high} of perpendicular viscous drag of the form $\Delta w \xi_{\perp}$, where $\Delta w$ is the relative perpendicular speed between the cylinder and the surrounding fluid and the coefficient $\xi_\perp$ is approximately constant throughout the cylinder. We define the function $\Lambda(d)$ as the ratio of the induced fluid speed due to the adjacent cylinder and the velocity of the adjacent cylinder. Thus, under the above assumptions, the deflection of the cylinders is governed by 
\begin{subequations}\label{PDE}
    \begin{align}
        s\frac{\partial^4 w_i}{\partial x^4}=-\xi_\perp \left[\frac{\partial w_i}{\partial t}-\Lambda\left(d=d_0+\frac{w_i-w_j}{j-i}\right)\frac{\partial w_j}{\partial t} \right], 
    \end{align}
supplemented by the boundary conditions
    \begin{align}
        \frac{\partial w_i(0,t)}{\partial x}=\phi_ie^{i2\pi(\omega_i t+(i-1)\gamma)},\quad w_i(0,t)=\frac{\partial^2 w_i(l,t)}{\partial x^2}=\frac{\partial^3 w_i(l,t)}{\partial x^3}= 0,\quad 
    \end{align}
\end{subequations}
where for cylinder $1$, $(i,j)=(1,2)$ and for cylinder $2$, $(i,j)=(2,1)$.

The flow field due to a slender cylinder moving relative to a viscous fluid in the creeping flow regime may be approximated by a uniform distribution of Stokeslets and dipoles positioned along the centerline of the cylinder. For motion perpendicular to the centerline, the magnitude of the Stokeslet distribution is $\xi_{\perp}\Delta w$, and the dipole magnitude is $r_c^2\xi_{\perp}\Delta w/4\mu$, where $\xi_\perp\approx 8\pi\mu/(0.386+\ln(l^2/r_c^2))$ \citep{lighthill1975mathematica}. Thus, the induced speed may be approximated \citep[using the same integral approximations and approach used in][]{lighthill1975mathematica} as
\begin{equation}\label{lamb}
    \Lambda \approx \frac{\xi_\perp}{4\pi\mu}\left[1+\ln\left(\frac{2l}{d_0+w_1-w_2}\right)\right].
\end{equation}

Eqs. (\ref{PDE}) may be decoupled by subtracting the equation governing cylinder $2$ from the equation governing cylinder $1$, and substituting relative deflection $w_d=(w_1-w_2)/2$, thus isolating $w_d$. Similarly, by addition of the governing equations of both cylinders and substituting average deflection $w_a=(w_1+w_2)/2$, the governing equation for $w_a$ may be obtained, which however does depends on $w_d$. We define the normalized axial coordinate $X=x/l$, normalized time $T=t2\pi\omega^*$, normalized average deflection $W_d=w_d/w^*_d$, normalized relative deflection $W_a=w_a/w^*_a$, normalized angular speeds $(\Omega_1,\Omega_2)=(\omega_1/\omega^*,\omega_2/\omega^*)$ and normalized gap $D=d/d_0$.

Substituting normalized variables, the equations governing $W_a$ and $W_d$ are
\begin{subequations}\label{PDE_norm}
    \begin{align}
        \frac{\partial^4 W_d}{\partial X^4}=-S_p^4 \left[1+\Lambda\left(D=1+2\varepsilon W_d\right) \right]\frac{\partial W_d}{\partial T} 
    \end{align}
    \begin{align}
        \frac{\partial^4 W_a}{\partial X^4}=-S_p^4 \left[1-\Lambda\left(D=1+2\varepsilon W_d\right) \right]\frac{\partial W_a}{\partial T}, 
    \end{align}
\end{subequations}
where $S_p=(\xi_\perp l^4 2\pi\omega^*/s)^{1/4}$ is the Sperm number. Eqs. (\ref{PDE_norm}) are supplemented by the normalized boundary conditions
\begin{subequations}\label{BC_total}
    \begin{align}
        W_d\big|_{X=0}=0,\quad\frac{\partial W_d}{\partial X}\Big|_{X=0}=\frac{\phi_1 l}{2w^*_d} e^{i\Omega_1 T}-\frac{\phi_2 l}{2w^*_d}e^{i(\Omega_2 T+2\pi\gamma)},\\ W_a\big|_{X=0}=0,\quad \frac{\partial W_a}{\partial X}\Big|_{X=0}=\frac{\phi_1 l}{2w^*_a} e^{i\Omega_1 T}+\frac{\phi_2 l}{2w^*_a}e^{i(\Omega_2 T+2\pi\gamma)},
    \end{align}
representing the hinge boundary condition and slope actuation at $X=0$ and 
    \begin{align}
        W_d\big|_{X=0}=\frac{\partial^2 W_d}{\partial X^2}\Big|_{X=1}=\frac{\partial^3 W_d}{\partial X^3}\Big|_{X=1}=W_a\big|_{X=0}=\frac{\partial^2 W_a}{\partial X^2}\Big|_{X=1}=\frac{\partial^3 W_a}{\partial X^3}\Big|_{X=1}=0
\end{align}
\end{subequations}
representing the free end at $X=1$. From (\ref{BC_total}), the characteristic values $w_a^*$ and $w_d^*$ may be defined as
\begin{subequations}\label{wdest}
    \begin{align}
        w_d^*=\max\left[\frac{\phi_1 l}{2} \cos{(\Omega_1 T)}-\frac{\phi_2 l}{2}\cos{(\Omega_2 T+2\pi\gamma)}\right]
\end{align}
and
    \begin{align}
    w_a^*=\max\left[\frac{\phi_1 l}{2} \cos{(\Omega_1 T)}+\frac{\phi_2 l}{2}\cos{(\Omega_2 T+2\pi\gamma)}\right].
\end{align}
\end{subequations}
Asymptotic expansion of the nonlinear (\ref{PDE_norm}) allows for approximation to a set of linear equations at the limit,
\begin{equation}\label{eps}
      \varepsilon=\frac{w_d^*}{d_0}\ll1,
\end{equation}
by presenting $\Lambda(D)$ as a Taylor series around $D=1$ 
\begin{equation}\label{lamb_taylor}
    \Lambda(D)\sim \Lambda(1)+\varepsilon 2W_d \frac{\partial \Lambda(D)}{\partial D}+\frac{(2\varepsilon W_d)^2}{2}\frac{\partial^2 \Lambda(D)}{\partial D^2},
\end{equation}
as well as asymptotically expanding $W_d$ and $W_a$
\begin{equation}\label{taylor_expand}
    W_d\sim W_{d,0}+\varepsilon W_{d,1}+\varepsilon^2 W_{d,2},\quad W_a \sim W_{a,0}+\varepsilon W_{a,1}+\varepsilon^2 W_{a,2}.
\end{equation}
Substituting (\ref{lamb_taylor}) and (\ref{taylor_expand}) into (\ref{PDE_norm}) and defining the differential operators $\textit{$L^+$}=\partial^4/\partial X^4+S_p^4(1+\Lambda(1))\partial/\partial T$ and $\textit{$L^-$}=\partial^4/\partial X^4+S_p^4(1-\Lambda(1))\partial/\partial T$, yields the leading-order $O(1)$ of (\ref{PDE_norm})
\begin{equation}\label{LeadingOrder}
    \textit{$L^+$}W_{d,0}=0, \quad \textit{$L^-$}W_{a,0}=0,
\end{equation}
as well as order $O(\varepsilon)$,
\begin{equation}\label{FirstOrder}
    \textit{$L^+$}W_{d,1}=-2S_p^4W_{d,0}\frac{\partial W_{d,0}}{\partial T}\frac{\partial \Lambda(D)}{\partial D}, \quad \textit{$L^-$}W_{a,1}=2S_p^4W_{d,0}\frac{\partial W_{a,0}}{\partial T}\frac{\partial \Lambda(D)}{\partial D},
\end{equation}
and order $O(\varepsilon^2)$,
\begin{equation}\label{SecondOrder}
    \begin{aligned}
    \textit{$L^+$}W_{d,2}=-2S_p^4\left[\frac{\partial W_{d,0}}{\partial T}\left(W_{d,1} \frac{\partial\Lambda (D)}{\partial D}+W_{d,0}^2\frac{\partial^2\Lambda (D)}{\partial D^2}\right)+\frac{\partial W_{d,1}}{\partial T}W_{d,0} \frac{\partial\Lambda (D)}{\partial D}\right],\\
    \textit{$L^-$}W_{a,2}=2S_p^4\left[\frac{\partial W_{a,0}}{\partial T}\left(W_{d,1} \frac{\partial\Lambda (D)}{\partial D}+W_{d,0}^2\frac{\partial^2\Lambda (D)}{\partial D^2}\right)+\frac{\partial W_{a,1}}{\partial T}W_{d,0}\frac{\partial \Lambda(D)}{\partial D}\right]
\end{aligned}
\end{equation}
and so forth. The boundary conditions for the $O(1)$ equations (\ref{LeadingOrder}) are identical to (\ref{BC_total}). For the $O(\varepsilon)$ (\ref{FirstOrder}) and $O(\varepsilon^2)$ (\ref{SecondOrder}) equations, the boundary conditions (\ref{BC_total}) are modified so that $\partial W_{d,1}/\partial X=\partial W_{a,1}/\partial X=\partial W_{d,2}/\partial X=\partial W_{a,2}/\partial X=0$ at $X=0$. 

The leading-order $W_{d,0}$ solution can be presented by
\begin{equation}\label{zero_sol}
    \begin{aligned}
        W_{d,0}&=\mathbb{Re}\left(e^{i\Omega_1T}F_{\Omega_1,d}(X)-e^{i\left(\Omega_2T+2\pi\gamma\right)}F_{\Omega_2,d}(X)\right)\\&=\frac{1}{2}\bigg[|F_{\Omega_1,d}(X)|\Big(e^{i\left(\Omega_1T+\angle F_{\Omega_1,d}(X) \right)}- e^{-i\left(\Omega_1T+\angle F_{\Omega_1,d}(X) \right)}\Big)\\&-|F_{\Omega_2,d}(X)|\Big(e^{i\left(\Omega_2T+2\pi\gamma+\angle F_{\Omega_2,d}(X) \right)}-e^{-i\left(\Omega_2T+2\pi\gamma+\angle F_{\Omega_2,d}(X) \right)}\Big)\bigg]
    \end{aligned}
\end{equation}
where the functions $F_{\Omega_1,d}(X)$,  $F_{\Omega_2,d}(X)$ are 
\begin{equation}\label{x_func_1}
    \begin{aligned}
        F_{\Omega_1,d}(X)=\frac{\phi_1 l}{2w^*_d}\frac{e^{i\frac{\pi}{8}}\left(S_p\sqrt[4]{\Omega_1(1+\Lambda(1))}\right)^{-1}}{(2+2\cos{\phi_{d}^1}\cosh{\phi_{d}^1})}(\sin{\theta_{d}^1}+\sinh{\theta_{d}^1}\\+\sin{\phi_{d}^1}\cosh{(\phi_{d}^1-\theta_{d}^1)}-\cos{\phi_{d}^1}\sinh{(\phi_{d}^1-\theta_{d}^1)}
        \\-\cosh{\phi_{d}^1}\sin{(\phi_{d}^1-\theta_{d}^1)}+\sinh{\phi_{d}^1}\cos{(\phi_{d}^1-\theta_{d}^1)})
    \end{aligned}
\end{equation}
and 
\begin{equation}\label{x_func_2}
    \begin{aligned}
        F_{\Omega_2,d}(X)=\frac{\phi_2 l}{2w^*_d}\frac{e^{i\frac{\pi}{8}}\left(S_p\sqrt[4]{\Omega_2(1+\Lambda(1))}\right)^{-1}}{(2+2\cos{\phi_{d}^2}\cosh{\phi_{d}^2})}(\sin{\theta_{d}^2}+\sinh{\theta_{d}^2}\\+\sin{\phi_{d}^2}\cosh{(\phi_{d}^2-\theta_{d}^2)}-\cos{\phi_{d}^2}\sinh{(\phi_{d}^2-\theta_{d}^2)}
  		\\-\cosh{\phi_{d}^2}\sin{(\phi_{d}^2-\theta_{d}^2)}+\sinh{\phi_{d}^2}\cos{(\phi_{d}^2-\theta_{d}^2)})
    \end{aligned}
\end{equation}
and where $\theta_{d}^i=XS_p\sqrt[4]{\Omega_i(1+\Lambda(1))}r_1$, $\phi_{d}^i=S_p\sqrt[4]{\Omega_i(1+\Lambda(1))}r_1$ and $r_1=\sqrt[4]{-i}=0.92-0.38i$.
Applying the relevant homogeneous boundary conditions and substituting (\ref{x_func_2}) into (\ref{FirstOrder}), we obtain that $W_{d,1}$ is of the form
\begin{equation}\label{first_sol}
    \begin{aligned}
        W_{d,1}&=\mathbb{Re}\bigg(F_{2\Omega_1,d}(X)e^{i2\Omega_1T}+F_{2\Omega_2,d}(X)e^{i2\Omega_2T}\\&+F_{(\Omega_1+\Omega_2),d}(X)e^{i(\Omega_1+\Omega_2)T}+F_{(\Omega_1-\Omega_2),d}(X)e^{i(\Omega_1-\Omega_2)T}\bigg)
    \end{aligned}
\end{equation}
where the functions $F_{2\Omega_1,d}(X)$, $F_{2\Omega_2,d}(X)$, $F_{(\Omega_1+\Omega_2),d}(X)$ and $F_{(\Omega_1-2\Omega_2),d}(X)$ may be readily obtained by substituting (\ref{zero_sol}) into (\ref{FirstOrder}), isolating each harmonic and solving the corresponding ordinary differential equation. Substituting (\ref{first_sol}) and (\ref{zero_sol}) into (\ref{SecondOrder}), yields $W_{d,2}$ with the harmonics $\Omega_1$, $3\Omega_1$, $\Omega_2$, $3\Omega_2$, $2\Omega_1+\Omega_2$, $\Omega_1+2\Omega_2$, $2\Omega_1-\Omega_2$ and $2\Omega_1-\Omega_2$. The corresponding functions $W_{a,0}$, $W_{a,1}$ and $W_{a,2}$ can be calculated by applying a similar approach and will have identical frequencies, but different mode functions, compared to $W_{d,0}$, $W_{d,1}$ and $W_{d,2}$ (see Appendix \ref{Appendix_W_a}).

\section{Deflection modes and Experimental Illustration}
\begin{figure}
    \begin{center}
        \includegraphics[width=1\textwidth]{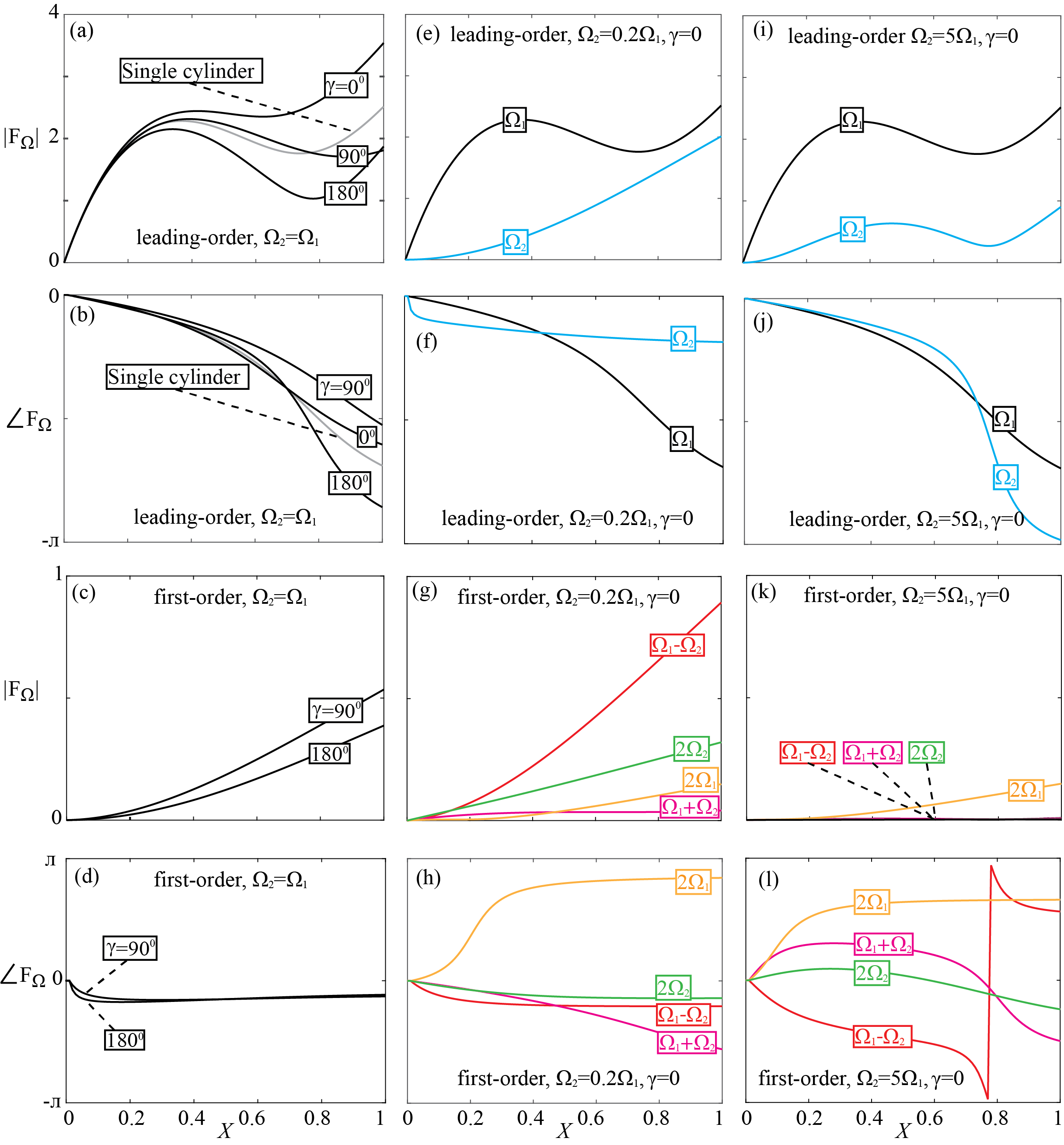}
        \caption{Axial distributions of the amplitude and phase of the frequencies comprising  $w_1$ for various configurations. In all cases cylinder $1$ is actuated with amplitude $\phi_1=15^0$ and frequency $\Omega_1=0.5$, the amplitude of cylinder 2 is $\phi_2=\phi_1$,  $\varepsilon=0.1$ and $S_p=2.1$. Panels (a-d) examine the effect of phase $\gamma$ difference for $\Omega_2=\Omega_1$. The modes presented in panels (a,b) oscillate at frequency $\Omega_1$ and the modes presented in panels (c,d) oscillate at frequency $2\Omega_1$. Panels (e-f) examine the effect of an adjacent oscillating cylinder with $\Omega_2=5\Omega_1$ and $\gamma=0$. Panels (i-l) examine  the effect of $\Omega_2=0.2\Omega_1$ and $\gamma=0$. }
        \label{fig:results}
    \end{center}
\end{figure}

Fig. \ref{fig:results} presents the amplitude and phase of the harmonics comprising  $w_1$ vs. the coordinate $X$, for various configurations where $\varepsilon=0.1$ and $S_p=2.1$. In all cases the cylinders are actuated at amplitudes $\phi_1=\phi_2=15^0$ and the normalized frequency  of cylinder 1 is $\Omega_1=0.5$. Panels (a-d) present the effect of phase difference $\gamma$ for cylinders actuated at identical frequencies $\Omega_2=\Omega_1$. For comparison, an isolated cylinder is presented by a grey smooth line. The interaction with an adjacent cylinder with $\gamma=0$  decreases the effective Sperm number and thus increases the deformation of the cylinder. This yields deflection dynamics identical to an isolated cylinder with a modified Sperm number $Sp(1-\Lambda(1))^{1/4}$, where $\Lambda(1)$ is the leading-order interaction term. Similarly, the leading-order effect of an adjacent cylinder oscillating at anti-phase $\gamma=\pi$ is to increase the effective Sperm number to $Sp(1+\Lambda(1))^{1/4}$, thus decreasing the deflection of the cylinder. For $\gamma=0$ the first-order is identically zero, while for both $\gamma=\pi/2$ and $\gamma=\pi$ (see panels c and d) the first-order correction is nearly independent of $X$ and includes only small value of phase $\approx10^0$. Panels (e-f) examine the effect of an adjacent oscillating cylinder with $\Omega_2=0.2\Omega_1$ and no phase $\gamma=0$. Similarly, panels (i-l) examine the opposite case of  $\Omega_2=5\Omega_1$ and $\gamma=0$. The leading-order effect of the adjacent cylinder is significant for $\Omega_2=0.2\Omega_1$ (where it has a similar effect to the leading-order direct actuation of the cylinder, see panel e). However, a much smaller effect is evident for the case $\Omega_2=5\Omega_1$ (see panel i). Panels (f) and (j) examine the phase of the leading-order modes, presenting an opposite effect where the phase is small and nearly uniform for $\Omega_2=0.2\Omega_1$ and significant for the case of $\Omega_2=5\Omega_1$. The first-order modes (see panels g and k) are dominated by the small frequencies, corresponding to smaller effective Sperm numbers of the interaction. Thus, for configurations in which the minimal frequency (from the set $2\Omega_1,2\Omega_2,\Omega_1+\Omega_2,|\Omega_1-\Omega_2|$) is significantly smaller than all other frequencies, the first-order dynamics may be reasonably approximated by the minimal frequency mode alone (see panel k). 


\begin{figure}
    \begin{center}
        \includegraphics[width=1\textwidth]{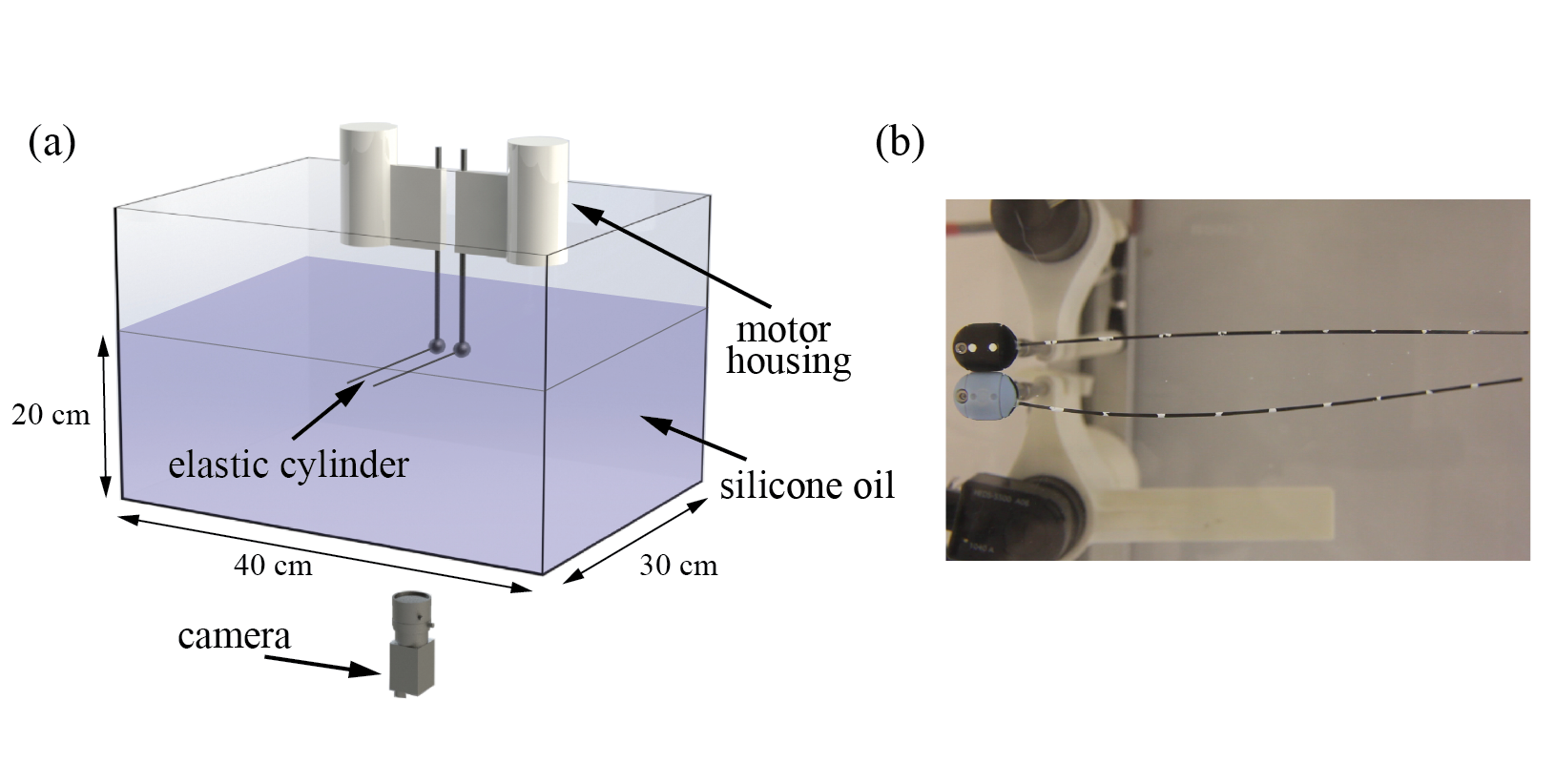}
        \caption{(a) Schematic description of the experimental setup consisting of two elastic cylinders deforming due to a prescribed oscillation of the slope at their bases. (b) An illustrative frame obtained by the Canon EOS 60D DSLR camera during an experiment. }
        \label{fig:exp_scheme}
    \end{center}
\end{figure}

Experiments were conducted to quantify the interaction between the two oscillating elastic cylinders and validate the results presented in \S2. The experimental setup is illustrated in Fig. \ref{fig:exp_scheme}. Actuation is achieved by a pair of Faulhaber 3257G024CR DC motors operating outside of the fluid, where motion is transferred to the cylinders through levers connected to elongated rotation axes. The bases of the cylinders are fixed while the slopes at the bases are forced to oscillate at predetermined amplitude and frequency. The motors controller is  iPOS4808 BX-CAN drive and the elastic cylinders are composed of carbon-fibre with diameter of $1\textrm{mm}$ and length of  $150\textrm{mm}$. The gap at rest between the cylinders is $d_0=14mm$. The immersing fluid is Xiameter\textregistered PMX-200 Silicone oil with viscosity $\mu=59.2 Pa\cdot s$ and density of $\rho_l=987 Kg/m^3$. The container dimensions are $0.4m\times0.3m\times0.2m$ and the elastic cylinders are placed symmetrically to both sides of the container center plane (see figure \ref{fig:exp_scheme}). A Canon EOS 60D DSLR camera with Canon EF-S $17-85\textrm{mm}$ f/4-5.6 IS USM lens was used to record the motion of both cylinders at 25 frames-per-second and resolution of $1920\; \times\; 1080$ pixels per frame. The recorded data was processed by open source code \citep{hedrick2008software}.

\begin{figure}
    \begin{center}
        \includegraphics[width=0.95\textwidth]{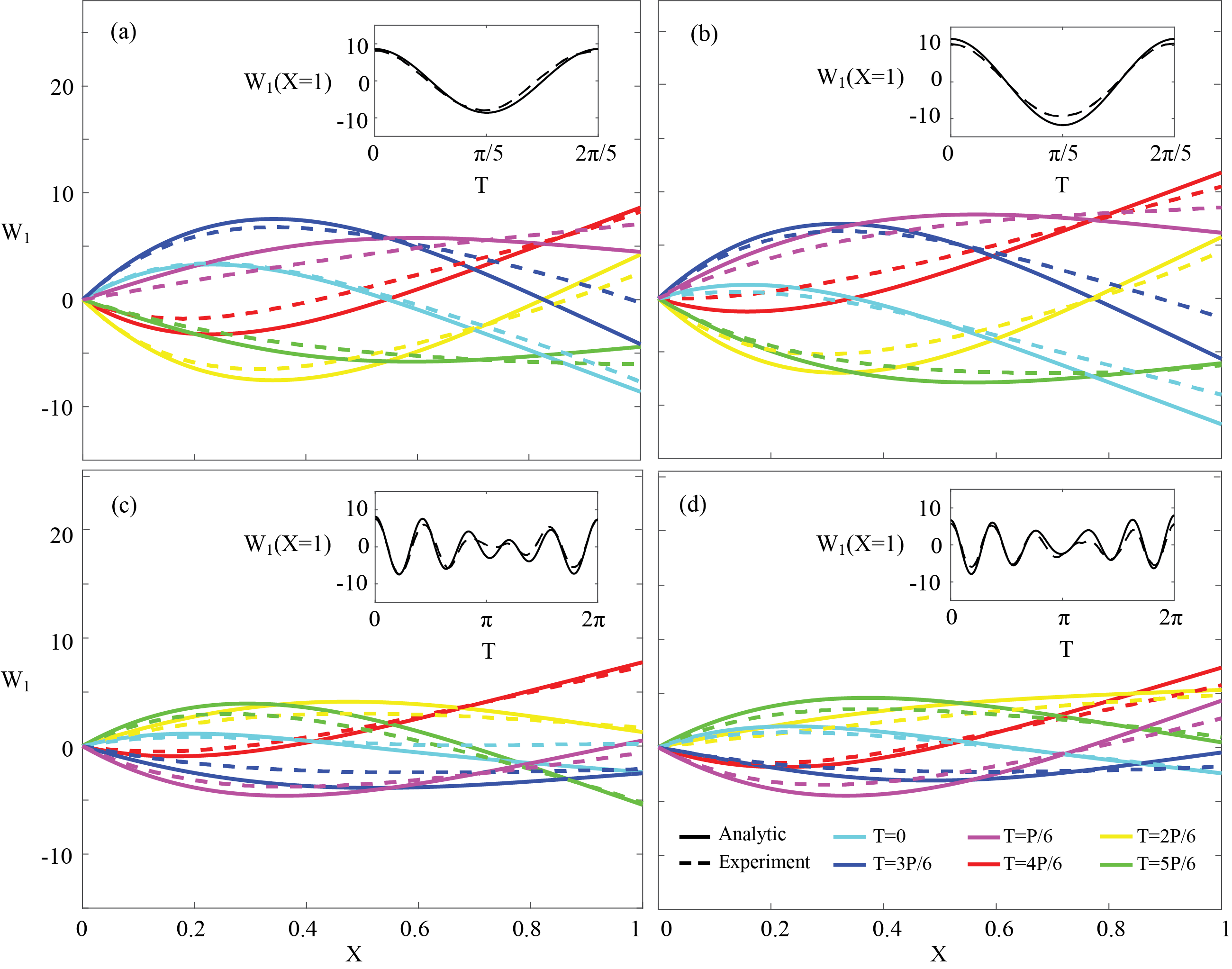}
        \caption{Deflection $W_1$ of cylinder 1  for $\Omega_1=5$ and amplitude $\phi_1=\phi_2=15^0$ at six equally spaced times along the full oscillation period $P$. Dimensional values are $w_1=W_1\times1.4mm$, $\omega=\Omega\times 0.1Hz$ and $t=T\times 0.62 s$. Smooth lines denote theoretical results and dashed lines denote experimental data. The examine cases are (a) without an adjacent cylinder, (b) an adjacent cylinder with $\Omega_1=\Omega_2$, (c) $\Omega_2=0.8\Omega_1$ and (d) $\Omega_2=1.2\Omega_1$.  Inserts present the deflection of the cylinder at its free end $W_1(X=1)$ vs. time. See supplementary information - movies 1-4.}
        \label{fig:tail_shape}
    \end{center}
\end{figure}

Fig. \ref{fig:tail_shape} presents the experimental (dashed lines) and theoretical (smooth lines) deflection patterns $W_1$ of cylinder $1$ oscillating at frequency $\Omega=0.5$ and amplitude at $\phi_1=15^0$ adjacent to cylinder $2$. The inserts present the deflection of the free end of the cylinder $W_1(X=1)$ for a full cycle period defined as $P$. Dimensional values are related to the normalized values by $w_1=W_1\times1.4mm$, $\omega=\Omega\times 0.1Hz$ and $t=T\times 0.62 s$. For reference, panel (a) (supplementary information - movie 1) presents the deflection of an elastic cylinder oscillating without the presence of a second adjacent cylinder. Panel (b) (supplementary information - movie 2) presents the deflection $W_1$ for the case of an adjacent cylinder oscillating at identical frequency $\Omega_1=\Omega_2$, identical amplitude $\phi_1=\phi_2$ and without phase $\gamma=0$. A significant increase in amplitude of the deflection is clearly evident and the deflection patterns remain symmetric in this case. Panels (c) and (d) (supplementary information - movies 3 and 4) present the effect of an adjacent cylinder oscillating at a slightly smaller frequency ($\Omega_2=0.8\Omega_1$ in panel c) and a slightly higher frequency  ($\Omega_2=1.2\Omega_1$ in panel d). Due to the multiple frequencies characterising panels (c) and (d), the full period is defined by the $\Omega_1-\Omega_2=1$ mode as $P=2\pi$, which is an integer multiplication of all other modes. The values of $\Omega_2$ in panels (c) and (d) were chosen to be similar to $\Omega_1$ in order to reduce the effective Sperm number of the $\Omega_1-\Omega_2$ mode, thus increasing the first-order deflection to be experimentally significant.

\begin{figure}
    \begin{center}
        \includegraphics[width=0.95\textwidth]{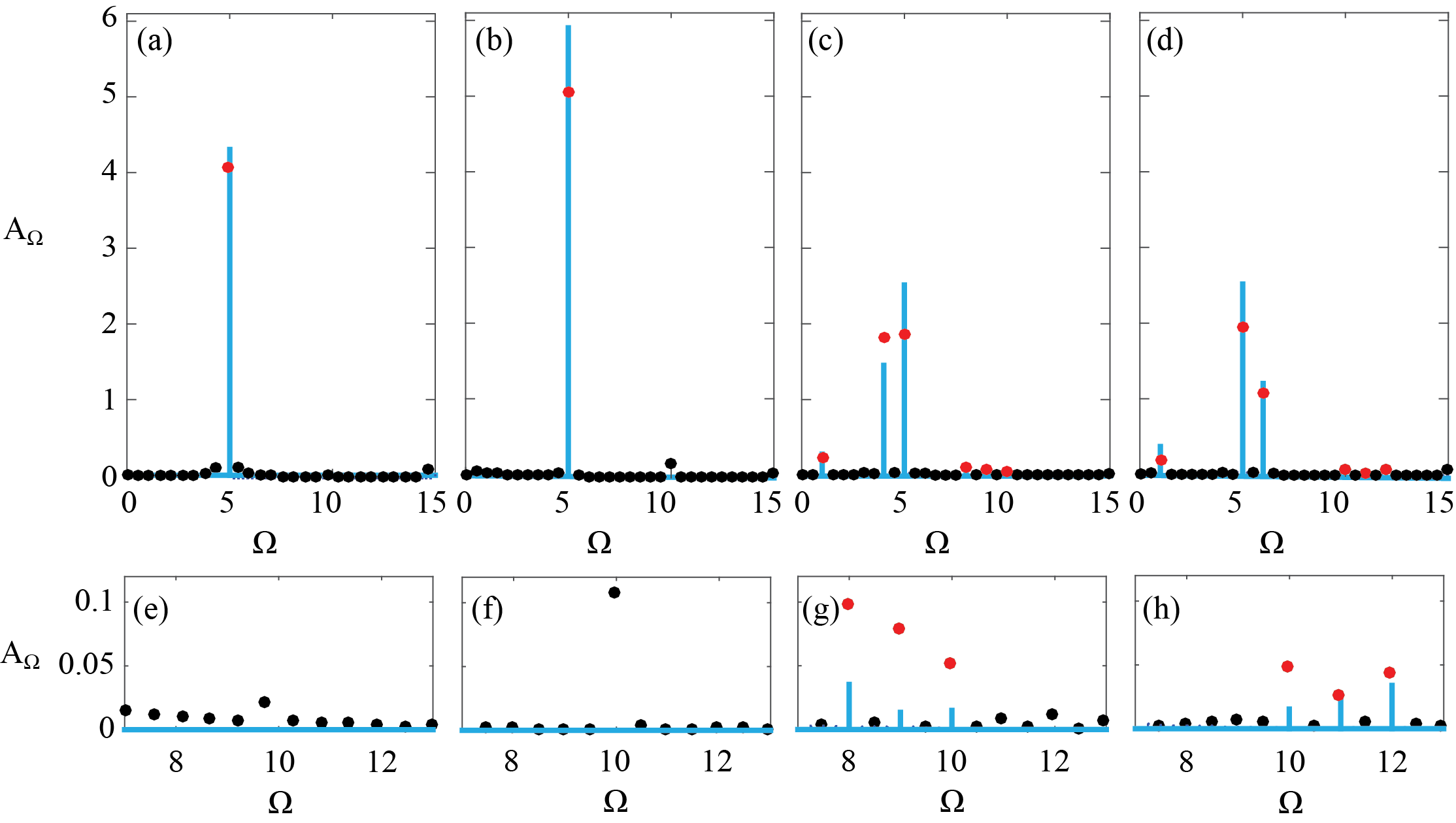}
        \caption{Fourier decomposition of the experimental deflection $W_1$ at ${X=1}$, presenting amplitude $A_\Omega$ vs. frequency $\Omega$. Panels (a-d) present the frequency decomposition of inserts of panels (a-d) in figure \ref{fig:tail_shape}, respectively. Filled circle markers denote experimental data and smooth blue lines denote the theoretical predictions. Red circles denote harmonics  expected from the theoretical results. Panels (e-h) are magnifications of (a-d) in order to present the amplitudes of the $2\Omega_1,2\Omega_2,\Omega_1+\Omega_2$ first-order frequencies.}
        \label{fig:fourier}
    \end{center}
\end{figure}

Fig. \ref{fig:fourier} presents frequency decomposition (based on MATLAB\textregistered FFT function) of the experimental data  presented in the inserts of Fig. \ref{fig:tail_shape}. The blue smooth lines are the theoretical predictions, and the full circles are the experimental amplitudes (predicted frequencies are filled red circles and other frequencies are filled black circles). As expected, since $\Omega_1\approx \Omega_2$ in panels (c,d), the deflection is dominated by the frequencies of the actuation in leading-order $\Omega_1$, $\Omega_2$ and the  $\Omega_1-\Omega_2$ first-order harmonic. All other frequencies are small compared with the experimental resolution. Panels (e-f) are magnifications of panels (a-d), focusing on of the amplitudes of the $2\Omega_1$, $2\Omega_1$, $\Omega_1+\Omega_2$ first-order frequencies. For all examined cases in figures and \ref{fig:tail_shape} and \ref{fig:fourier}, a reasonable agreement between the experimental data and the theoretical results is evident.



\section{Concluding Remarks}

While microscopic swimmers do not create propulsion by oscillating  the slope at the base of their flagella, the results may still provide insight for biological mechanisms which inherently involve elasticity. The presented analysis yielded that the effect of an adjacent oscillating cylinder, with identical in-phase actuation frequency and amplitude, is to decrease the effective Sperm number. Thus, the optimal propulsion oscillation frequency for an array of flagella may be expected to be greater compared with the optimal frequency of an isolated flagellum. For $\Omega_1\approx \Omega_2$, the slowest $\Omega_1-\Omega_2$ mode dominates the first-order dynamics since the amplitude of deflection is inverse to the effective Sperm number and the mode's frequency. Future work may examine non-linear effects, propulsion dynamics, internal actuation distributed along the cylinder, as well as a study of the dynamics of a lattice of oscillating cylinders.



\appendix
\section{leading-order average deflection, $W_{a,0}$}\label{Appendix_W_a}
\begin{equation}\label{zero_sol_a}
    \begin{aligned}
        W_{a,0}&=\mathbb{Re}\left(e^{i\Omega_1T}F_{\Omega_1,a}(X)+e^{i\left(\Omega_2T+2\pi\gamma\right)}F_{\Omega_2,a}(X)\right)\\&=\frac{1}{2}\bigg[|F_{\Omega_1,a}(X)|\Big(e^{i\left(\Omega_1T+\angle F_{\Omega_1,a}(X) \right)}- e^{-i\left(\Omega_1T+\angle F_{\Omega_1,a}(X) \right)}\Big)\\&+|F_{\Omega_2,a}(X)|\Big(e^{i\left(\Omega_2T+2\pi\gamma+\angle F_{\Omega_2,a}(X) \right)}-e^{-i\left(\Omega_2T+2\pi\gamma+\angle F_{\Omega_2,a}(X) \right)}\Big)\bigg]
    \end{aligned}
\end{equation}
\begin{equation}\label{x_func_1_a}
    \begin{aligned}
        F_{\Omega_1,a}(X)=\frac{\phi_1 l}{2w^*_a}\frac{e^{i\frac{\pi}{8}}\left(S_p\sqrt[4]{\Omega_1(1-\Lambda(1))}\right)^{-1}}{(2+2\cos{\phi_{a}^1}\cosh{\phi_{a}^1})}(\sin{\theta_{a}^1}+\sinh{\theta_{a}^1}\\+\sin{\phi_{a}^1}\cosh{(\phi_{a}^1-\theta_{a}^1)}-\cos{\phi_{a}^1}\sinh{(\phi_{a}^1-\theta_{a}^1)}
        \\-\cosh{\phi_{a}^1}\sin{(\phi_{a}^1-\theta_{a}^1)}+\sinh{\phi_{a}^1}\cos{(\phi_{a}^1-\theta_{a}^1)}),
    \end{aligned}
\end{equation}
\begin{equation}\label{x_func_2_a}
    \begin{aligned}
        F_{\Omega_2,a}(X)=\frac{\phi_2 l}{2w^*_a}\frac{e^{i\frac{\pi}{8}}\left(S_p\sqrt[4]{\Omega_2(1-\Lambda(1))}\right)^{-1}}{(2+2\cos{\phi_{a}^2}\cosh{\phi_{a}^2})}(\sin{\theta_{a}^2}+\sinh{\theta_{a}^2}\\+\sin{\phi_{a}^2}\cosh{(\phi_{a}^2-\theta_{a}^2)}-\cos{\phi_{a}^2}\sinh{(\phi_{a}^2-\theta_{a}^2)}
  		\\-\cosh{\phi_{a}^2}\sin{(\phi_{a}^2-\theta_{a}^2)}+\sinh{\phi_{a}^2}\cos{(\phi_{a}^2-\theta_{a}^2)}).
    \end{aligned}
\end{equation}
and where $\theta_{a}^i=XS_p\sqrt[4]{\Omega_i(1-\Lambda(1))}r_1$, $\phi_{a}^i=S_p\sqrt[4]{\Omega_i(1-\Lambda(1))}r_1$ and $r_1=\sqrt[4]{-i}$.


\bibliographystyle{jfm}

\bibliography{interaction}

\end{document}